\documentstyle[multicol,aps,pra,epsf,prabib]{revtex}

\begin{document}
\title{Chord distribution functions of three-dimensional random media:
Approximate first-passage times of
Gaussian processes\footnote{Submitted to Physical Review E}}
\author{A. P. Roberts\footnote{Permanent address: Centre for
microscopy and microanalysis, University of Queensland, Brisbane,
Queensland 4072, Australia}}
\address{Princeton Materials Institute and Department of
Civil Engineering and Operations Research, \\
Princeton University, Princeton, New Jersey 08544, U.S.A.}
\author{S. Torquato\footnote{On sabbatical leave. Permanent address:
Princeton Materials Institute and Department of
Civil Engineering and Operations Research,
Princeton University, Princeton, New Jersey 08544, U.S.A.}}
\address{School of Mathematics, Institute for Advanced Study,
New Jersey 08540, U.S.A.}
\date{26 January 1999}
\maketitle
\begin{abstract}
The main result of this paper is a semi-analytic approximation
for the chord distribution functions of three-dimensional models
of microstructure derived from Gaussian random fields.
In the simplest
case the chord functions are equivalent to a standard
first-passage time problem, i.e., the probability
density governing the time taken by a Gaussian random process
to first exceed a threshold.
We obtain an approximation based on the assumption that
successive chords are independent. The result is a generalization
of the independent interval approximation recently used to
determine the exponent of persistence time decay in
coarsening.
The approximation is easily extended to more general
models based on the intersection and union sets of
models generated from the iso-surfaces of random fields.
The chord distribution functions play an important role
in the characterization of random composite and porous materials. 
Our results are compared with experimental data obtained
from a three-dimensional image of a porous Fontainebleau
sandstone and a two-dimensional image of a
tungsten-silver composite alloy.

\noindent
PACS numbers(s): 02.50.-r, 05.40.+j, 81.05.Rm, 47.55.Mh
\end{abstract}

\begin{multicols}{2}
\section{Introduction}
The statistical characterization and modeling of two-phase
disordered microstructure is a central problem in many
fundamental and applied sciences~\cite{TorqRev91}. 
Predicting the properties of disordered materials
relies on the availability of accurate microstructural models,
which rely in turn on accurate statistical characterization.
After the volume fraction of each phase, and interfacial
surface area, the most important statistical quantity is the two-point
correlation function which is obtained from cross-sectional
micrographs, or small angle scattering experiments.
Although the two-point correlation function is very
useful, there are a variety of important problems
where more detailed statistical information is necessary. 
Another useful characteristic of microstructure, which
has proved essential in theory and application, is
the chord-length distribution
function~\cite{Serrabook}.  The chord functions play an
important role in stereology~\cite{UnderwoodBook},
mineralogy~\cite{KingBool},
the interpretation of small angle x-ray scattering data~\cite{Levitz92},
and have been incorporated in theories of mass
transport in porous media~\cite{TokGas}. Recently the chord
functions (and the
related ``lineal-path'' function~\cite{TorqLu93}) have been
employed in the generation of three-dimensional (3D)
microstructural models for predicting macroscopic
properties~\cite{RobertsRec,BourgeoisFilter,YeongRec2}.
In this paper, we derive approximate forms of the chord functions
for a relatively new model of random media based on Gaussian random
fields.

\begin{figure}
{\samepage\columnwidth20.5pc
\centering \epsfxsize=8.0cm\epsfbox{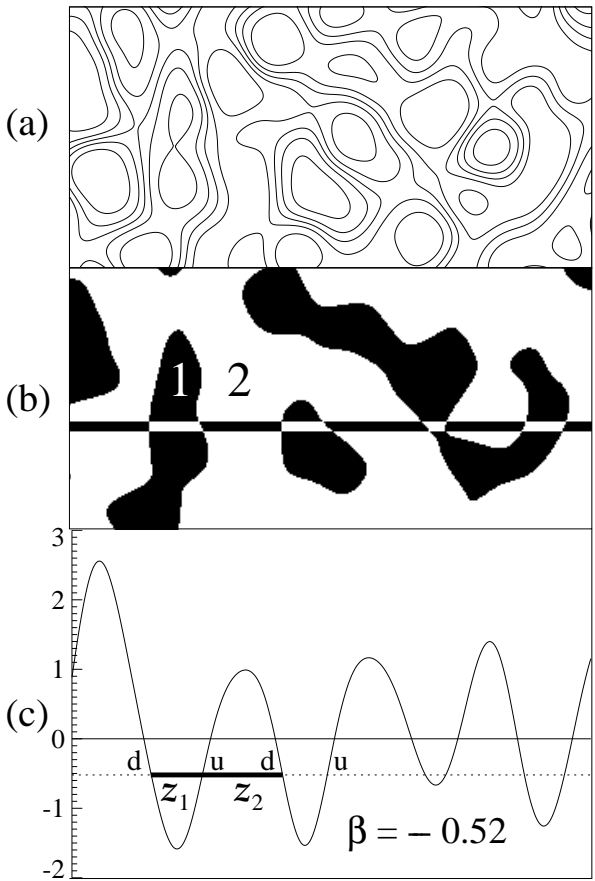}
\caption{Generation of a two-phase model (b) by thresholding
a Gaussian random field (a). The chord length distribution
functions are calculated by counting the number of
chords of a given length (b). In one dimension the
chord lengths $z_1$ and $z_2$ are defined by up- and down-crossings
(shown as $u$ and $d$ respectively) of a threshold $\beta$ by
a random process (c).  \label{cdf}}
}
\end{figure}

A useful model of two-phase random porous and composite media
is obtained by modeling the internal interface
of the microstructure as the iso-surface (or level-cut) of a
correlated Gaussian random field
$y({\bf r})$~\cite{Hopper82,Quiblier84,Berk91,Blumenfeld93,Roberts95a}.
A region of space can be divided into two phases
(e.g.\ pore and solid) according
to whether $y({\bf r})$ is less or greater than some threshold.
We define phase 1 to occupy the region where
$y({\bf r})\leq\beta$ and phase 2
to occupy the region where $y({\bf r})>\beta$.
This is illustrated for two dimensions in  Figs.~\ref{cdf}(a) and (b). 
The model, and variants, have proved useful in describing the
microstructure of many different
materials~\cite{RobertsRec,Adlerbook,LevitzRec}; and a more
thorough characterization of the microstructure is an important goal.
The two- and three-point correlation functions of the model
can be calculated, but the chord functions are presently measured
from simulations~\cite{RobertsRec,LevitzRec}.
There are advantages (for speed, accuracy
and interpretation) in obtaining analytic expressions for the
chord functions of the model. As we show, the problem is
equivalent to finding the probability density governing the
time it takes for a Gaussian random process to first cross an arbitrary
threshold. This is a conventional first-passage time problem.

First-passage time problems arise in
many branches of physics~\cite{ZeitakExp}, information
theory~\cite{Rainal88}, queuing theory~\cite{Hallbook},
ocean science~\cite{PodgorUP} and
reliability studies in the engineering
sciences~\cite{Nielsen90} amongst others.
For this reason the problem has received
a great deal of attention. Rice actually provided a formal series 
solution to the problem for Gaussian processes~\cite{RicePR}.
However, the series involves very difficult integrals of which
only the first is
generally evaluated~\cite{ZeitakExp};
the results being accurate for small time.
There are many approaches to finding useful approximations
for first-passage times (the aforementioned references provide
reviews of the literature in each field).
Here we restrict attention to methods based on the assumption
that the lengths of successive chords are independent. This idea
can be traced back to Siegert~\cite{Siegert51}
and McFadden~\cite{McFadden56}.
The approximation we use is most clearly derived from the
independent interval process~\cite{Levitz92,Mering68} for which the
assumption of independent chords is true by definition. This approach
was recently suggested for the calculation of persistence times
in coarsening~\cite{DerridaPers,MajumdarPers}.
The method can be extended to obtain the chord-distributions,
and actually provides an extremely useful way of viewing related problems
in different fields. 
In the following sections we derive some properties of
the chord function and the independent interval process.
The model is then applied to approximate the chord functions
of level-cut Gaussian random fields.  Finally, we compare the theoretical
results with experimental data.

\section{Chord distribution functions}

For a two-phase medium, there is a chord-distribution associated with
each phase $p_{i}(z)$ ($i$=1,2). The quantity $p_i(z)dz$ is defined as the
probability that a randomly chosen chord in phase
$i$ (a line segment with end points on the phase interface)
has length in the range $[z,z+dz]$.
In Fig.~\ref{cdf}(b), we illustrate calculation of $p_1(z)$ and
$p_2(z)$ for isotropic media: An ``infinite'' line (or many shorter ones
with arbitrary orientation) is drawn through the material,
and the number of chords in phase 1 [$N_1(z)$] with length in
the range $[z,z+dz]$ is counted.
If $N$ is the total number of chords of phase 1, then
$p_1(z)dz= N_1(z)/N$. The quantity $p_2$ is defined in an analogous manner.
An important quantity is the number of phase crossings per unit
length $n_c=2N/L$.  A fundamental relation in stereology gives
$n_c=\frac12s_v$ where $s_v$ is the specific surface, i.e., the surface
area to total volume ratio of a 3D composite.
Another useful relation follows from the fact that total length
of the chords is equal to that of the original line:
$ \sum_z N_1(z) z + \sum_z N_2(z) z =L$. If we divide by $N$ and convert
the sums to integrals we have
$\int p_1(z) z dz + \int p_1(z) z dz = 2/{n_c}$.
Similarly $ \sum_z N_i(z) z =\phi_i L$
where $\phi_i$ is the volume fraction of phase $i$. This gives the relation
$\langle z_i \rangle=\int p_i(z) z dz  = 2\phi_i/n_c$.
These equalities are used extensively below.
A useful statistical quantity is the
lineal-path function of each phase $L_i(z)$ ($i=1,2$), which
represents the probability that a random line segment
of length $z$ thrown into the material falls completely within
phase $i$ and is related to $p_i(z)$ by
$L_i(z)=\frac 12 n_c \int_z^\infty p_i (x) (x-z) dx$~\cite{TorqLu93}.

One reason for the usefulness of the chord function
is that $p_i(z)$ can be directly interpreted
in terms of observable microstructure features. First, if $p_i(0)\ne 0$ then,
at the resolution at which it is measured, phase $i$ contains
sharp corners. Second, the value at which $p_i(z)$ takes a maximum value
provides an estimate of the length scale associated with phase $i$.
Another is provided by the average chord length $\langle z_i \rangle$.
Third, if $p_i(z)\neq 0$ for large $z$, connected regions in phase $i$ at
scale $z$ exist. This
``connectedness'' information (along a lineal path) is clearly
important if long-range
phenomena (like macroscopic properties) are to be studied.
The direct relationship between $p_i(z)$ and
morphology shows that the chord functions give
a strong ``signature'' of microstructure and are
therefore an important tool in the characterization of
porous and composite media.

The chord and lineal-path functions are
closely related to first-passage times in the theory of stationary
time-dependent random processes. The analogy is
shown in Fig.~\ref{cdf}(c). If the values
of a random-field along a line are plotted against time
a one-dimensional (1D) random process $y(t)$ is obtained. If
a down-crossing of the threshold $\beta$ occurs at $t=t_d$
[so $y(t_d)=\beta$], the probability that the process
first exceeds the threshold in the interval $[t_d+t,t_d+t+dt]$ is
$p_1(t)dt$. The corresponding density for the first down-crossing
after an up-crossing (sometimes called
the second-passage time) is $p_2(t)$.
There are several other common first-passage
times which relate to our work.
Suppose $y(t)$ is a random process representing the
response of an electrical or mechanical component which
fails if $y(t)$ exceeds some (generally high) threshold ($y(t)>\beta$).
A key quantity is the probability of failure $F(t)$
in the interval $[0,t]$~\cite{Vanmarcke75}.  Since the safe region is 
phase 1 the failure probability is just $F(t)=1-L_1(t)$
[$F(0)=1-L_1(0)=\phi_2$ being the probability of instantaneous failure].
A related quantity is the probability distribution of failure
times $f_1(t)$ given that $t=0$ is in the safe
region~\cite{Nielsen90}. This is given by
$f_1(t)=-L_1'(t)/\phi_1$.

\section{Independent Interval Process}

The independent interval process~\cite{Levitz92,Mering68}
is constructed by taking the
lengths of successive intervals to be independent of one
another and distributed according to the probability (chord) distribution
functions $p_1(z)$ and $p_2(z)$ for phase 1
and 2 respectively (this implies a stationary process).
Since this completely defines
the process it is possible to derive all other statistical
properties (such as the two-point correlation function) from
$p_1(z)$ and $p_2(z)$.
The indicator function $ I({\bf x})$ (which is unity in phase 1
and zero in phase 2) is very useful in
this regard. From above we have 
\begin{eqnarray}
n_c&=&\langle | I'({\bf x}) | \rangle=\frac{2}{ \langle z_1 \rangle +
\langle z_2 \rangle}  \\
\phi_1&=&\langle I({\bf x}) \rangle= \frac12 n_c \langle z_1  \rangle \\
\phi_2&=&\langle 1- I({\bf x}) \rangle =\frac12 n_c \langle z_2  \rangle.
\end{eqnarray}
Note that the results are true irrespective of correlations between
successive chords.

Now consider the two-point correlation
function which is defined as
\begin{equation}
S_{11}(x)=\langle I({\bf x}_1)I({\bf x}_2) \rangle.
\end{equation}
Stationarity of the process (and isotropy in more than one
dimension) implies that the average
only depends on $x=|{\bf x}_2-{\bf x}_1|$. 
To express $S_{11}$ as a function of $p_1$ and $p_2$ it is necessary to
derive some preliminary results.
The method is adapted from Ref.~\cite{DerridaPers}. 
Although the model is independent of any random field it is useful
to describe the left-hand ends of chords in phase 1 and 2
as down-crossings and up-crossings, respectively (See Fig~\ref{cdf}).
Consider the probability $R_{u1}(x)$ that a point at
distance $x$ to the right of an up-crossing (or right-hand end of
a chord in phase 1) falls in phase 1. Suppose
the chord (of phase 2) immediately to the right of the
down-crossing has length $z_2$. If $z_2>x$ then the point
falls in phase 2. If $z_2<x$ the
chance that the point falls in phase 1 is $R_{d1}(x-z_2)$,
where $R_{d1}(y)$ is the probability that that a point
a distance $y$ to the right of a down-crossing falls
in phase 1.
Thus
\begin{equation}
R_{u1}(x)=\int_0^x dz_2 p_2(z_2) R_{d1}(x-z_2).
\end{equation}
Now consider the converse problem for $R_{d1}(x)$.
Let $z_1$ be the length of the first chord (of phase 1) to the
right of a down-crossing.  The probability that a point a distance
$x$ from the down-crossing falls in phase 1 is unity
if $z_1 > x$ and $R_{u1}(x-z_1)$ if $z_1 < x$ giving
\begin{equation}
R_{d1}(x)=\int_x^\infty\!\!\! dz_1 \; p_1(z_1) + \int_0^x\!\!\! dz_1 \; p_1(z_1)
R_{u1}(x-z_1).
\end{equation}
Taking Laplace transforms of both equations and solving 
we have 
\begin{eqnarray}
\widehat{R_{u1}}&=& \frac{ \hat p_2 (1- \hat p_1) }{s(1-\hat p_1\hat p_2)}\\
\widehat{R_{d1}}&=& \frac{ (1- \hat p_1) }{s(1-\hat p_1\hat p_2)}
\end{eqnarray}
where $\hat f$=$\hat f(s)$=$\int_0^\infty \! e^{-sx}f(x) dx$
denotes the usual Laplace transform.

Next we need the probability $Q_1(y)\delta y$ that an arbitrary
point will fall in phase 1 a distance $[y,y+\delta y]$ from the
left of the first down-crossing on its right. The chance that
the point falls on a 1-chord of length $[z_1,z_1+dz_1]$ is  
$\frac12 n_c p_1(z_1) z_1 dz_1$. The point will be uniformly distributed
along the chord, therefore if it falls on a chord of length
$z_1$ the probability that it lies a distance $[y,y+\delta y]$ from
the left end is just $\delta y/z_1$. To obtain the
total probability we must sum over all chords with $z>y$ giving,
\begin{eqnarray}
Q_1(y)\delta y & = & \frac{n_c}2 \int_y^\infty \!\! dz_1 p_1(z_1)z_1
\frac {\delta y}{z_1} \\ 
\hat Q_1&=&\frac{n_c}{2s}(1-\hat p_1).
\end{eqnarray}
Note that $Q_1(0)=\frac 12 n_c$ as it should since the probability that
a point lies within a distance $\delta y$ (to the left) of
a down-crossing must be $\frac 12 n_c \delta y$. 

Recall that $S_{11}(x)$ is the probability that two points a
distance $x$ apart fall in phase 1.  
The chance that the second point falls in phase 1 (given that
the first does) depends on whether
the distance to the first up-crossing $y$ is greater or less than
the distance $x$.  If $y>x$ then
the right-hand point falls in phase 1 with unit probability.
If $y<x$ then the right-hand point falls in phase
1 with probability $R_{u1}(x-y)$. Thus we have
\begin{equation}
S_{11}(x)=\int_x^\infty\! dy Q_1(y) + \int_0^x\! dy Q_1(y) R_{u1}(x-y).
\end{equation}
Note that $S_{11}(0)=\phi_1$ as it should
since $S_{11}(0)$ is just the probability that a single point falls
in phase 1.  Taking Laplace transforms we obtain 
\begin{equation}
\hat S_{11} = \frac{\phi_1}s -
\frac{n_c}{2s^2}\frac{(1-\hat p_1)(1-\hat p_2)}{1-\hat p_1\hat p_2}.
\label{s11hat}
\end{equation}

Two other correlation functions of the independent interval process
are important for later discussions. The first,
$S_{c1}(x)=\langle |I'({\bf x}_1)||I({\bf x}_2)|\rangle$ is the 1D
analog of the surface-void correlation function which arises
in the study of 3D porous materials~\cite{TorqRev91}.
The second is the crossing-crossing correlation function
$S_{cc}(x)=\langle |I'({\bf x}_1)||I'({\bf x}_2)|\rangle$ analogous
to the surface-surface correlation function~\cite{TorqRev91}.
Again, stationarity implies that $S_{c1}$ and $S_{cc}$ depend only
on the distance $x=|{\bf x}_2-{\bf x}_1|$.

From the definition of $S_{c1}$ we have
\begin{equation}
\epsilon S_{c1}(x)  \approx 
\left\langle \left|I({\bf x}_1+\frac\epsilon2)-I({\bf x}_1-\frac\epsilon2)\right|
I({\bf x}_2) \right\rangle 
\end{equation}
where $\epsilon$ is small.
The expression on the right-hand side is the probability that
$x_1$ lies within a distance $\frac\epsilon2$ (which we call an
$\epsilon$-interval) of a crossing and that $x_2$ lies in phase 1.
Without loss of generality, we consider the case $x_2>x_1$.
Now there is an equal chance of the first point landing
in an $\epsilon$-interval of an up or down
crossing (probability $\frac12\epsilon n_c$). The
probability that $x_2$ falls in phase 1 is then either
$R_{u1}(x)$ or $R_{d1}(x)$. Hence, for $\epsilon\to0$,
\begin{equation}
\epsilon S_{c1}(x) =
\frac12\epsilon n_c \times \left[ R_{u1}(x) + R_{d1}(x) \right]
\end{equation}
or, using Laplace transforms,
\begin{equation}
\hat S_{c1} = \frac{n_c}{2s}\frac{(1-\hat p_1)(1+\hat p_2)}{1-\hat p_1\hat p_2}.
\label{sc1hat}
\end{equation}

Similarly we write the crossing-crossing correlation function as
\begin{eqnarray}
\epsilon^2 S_{cc}(x) \nonumber
& \approx&
\left\langle \left|I({\bf x}_1+\frac12\epsilon)-I({\bf x}_1-\frac12\epsilon)\right|
\right.  \\
&& \times \left. \left|I({\bf x}_2+\frac12\epsilon)-
I({\bf x}_2-\frac12\epsilon)\right| \right\rangle 
\end{eqnarray}
where the expression on the right is clearly the probability that
both points lie within an $\epsilon$-interval of a crossing.
To express this quantity in terms of the chord-distributions several
additional functions are needed. Let $R_{ud}(x)$ be the probability
that a point a distance $x$ from an up-crossing falls in an
$\epsilon$ interval of a down crossing. This can occur in two
ways; either the first chord of phase 2 adjacent the up-crossing
has length $z_2=x$, or $z_2<x$ in which case the point falls
near a down crossing with probability $R_{dd}(x-z_2)$, where $R_{dd}(y)$
is the  probability
that a point a distance $y$ to the right of a down-crossing is
in an $\epsilon$-interval of a down-crossing.
This is expressed as
\begin{equation}
R_{ud}(x)=\epsilon p_2(x) + \int_0^{x} dz_2 p_2(z_2) R_{dd}(x-z_2)
\label{mcfadden}
\end{equation}
The functions $R_{uu}$, $R_{du}$ are similarly defined and three additional
relations amongst the four functions can be derived and solved
to give
\begin{eqnarray}
\hat R_{ud}&=&{\epsilon \hat p_2}/(1- \hat p_1\hat p_2) \\
\hat R_{du}&=&\epsilon {\hat p_1}/(1- \hat p_1\hat p_2)  \\
 R_{dd}=R_{uu}&=& {\epsilon \hat p_1 \hat p_2}/(1- \hat p_1\hat p_2). 
\end{eqnarray}
Now from the definition of $S_{cc}$ it is clear that
\begin{equation}
\epsilon^2 S_{cc}(x)
 = \frac12 \epsilon n_c \left[ R_{uu}+R_{ud}+R_{du}+R_{dd} \right]
\end{equation} as $\epsilon\to0$. After taking Laplace Transforms
this gives
\begin{equation}
\hat S_{cc}=\frac{n_c}{2}\frac{\hat p_1+\hat p_2+2\hat p_1\hat p_2}
{1-\hat p_1\hat p_2}.
\label{scchat}
\end{equation}
A final expression which is useful for relating this work
to prior approaches is
\begin{equation}
R_{1u}(x)=\epsilon f_1(x) + \int_0^x \! dy R_{uu} (x-y) f_1(y),
\label{MK}
\end{equation}
where $R_{1u}(x)$ is the probability that a point a distance $x$ from
a point in phase 1 falls within an $\epsilon$-interval of an up-crossing,
and $f_1(x)$ is the probability distribution of ``failure times'' discussed
earlier.

Although this analysis of the independent interval process,
and the results for $S_{c1}$ and $S_{cc}$ appear new,
variants of Eq.~(\ref{s11hat}) for $S_{11}$ have
actually been derived in several quite different contexts. It is
useful to briefly demonstrate these connections.
The Poisson-Boolean model~\cite{Serrabook} of random two-phase media
has been widely studied and applied. In this model,
grains (which may be of different shapes and sizes) are placed
at uncorrelated random points in space (so the grains may overlap).
For spherical inclusions this
it is just the overlapping sphere (or Swiss-cheese) model~\cite{TorqRev91}.
For convex grains of any shape or size, the
chord-distribution function of the phase
exterior to the grains is $p_1(z)= \lambda \exp(-\lambda z)$ where
$\lambda=s_v/4\phi_1$.
Since the grains are uncorrelated in space
and convex it is clear that the length of successive chords along
any line through the a realization of the model will be independent.
Therefore Eq.~(\ref{s11hat}) applies. Specializing to Boolean
models [$\hat p_1 = \lambda/( \lambda +s)$]
we recover the well known result (e.g.~Ref.~\cite{KingBool})
\begin{equation}
\frac{s \hat S_{11}}{\phi_1}=\frac{1-\hat p_1}{1-\hat p_1\hat p_2}.
\end{equation}
Since $S_{11}(x)$ is  known for many Boolean models,
this result allows $p_2$ to be calculated;
\begin{equation}
\hat p_2 =1 +s/\lambda - \phi_1 / \lambda \hat S_{11}. 
\end{equation}
Let us further specialize to the case of a 1D
Boolean process where the grains are rods with random lengths
distributed according
to the cumulative distribution function $\Psi(z)$. For this model 
$S_{11}(x)=\phi_1 \exp\left( -\lambda \int_0^x (1-\Psi(z))dz \right)$
and the formula for $\hat p_2$ becomes a well known result for the busy
period in an M/G/$\infty$ queue~\cite{Hallbook,QuinTorq96a}.

The independent interval process is useful in the interpretation of
small-angle x-ray scattering data.
In Refs.~\cite{Levitz92,Mering68} the Fourier transform of
$d^2S_{11}/dx^2$ was derived in terms of the Fourier transforms of
the chord functions.  The result can be shown to be equivalent
to Eq.~(\ref{s11hat}).  Rice's~\cite{RicePR} formulae for $S_{11}$ for
two different types of random telegraph signal can also be
re-derived using Eq.~(\ref{s11hat}).  
In the next section we show how the independent interval process
can be used to derive useful approximations of the
chord functions for the single level-cut GRF model.

\section{Chord functions of level-cut Gaussian random fields}

A Gaussian field is statistically specified in terms
of a field-field correlation function $\langle y({\bf r}_1)y({\bf r}_2) 
\rangle=g(|\bf{r_2}-\bf{r_1}|)$~\cite{RicePR,RJAdler_book}.
Here we consider isotropic stationary fields 
with zero mean [$\langle y({\bf r}) \rangle=0$] and unit
variance [$ \langle y^2({\bf r}) \rangle =g(0)=1$].
Many methods exist for generating GRF's with a given $g(r)$.
For example, in 1D we have
\begin{equation}
y(x)=\sum_{i=1}^N a_i \cos k_i x + b_i \sin k_i x;\;\; k_i=\frac{2\pi i}T
\label{y1D}
\end{equation}
where $a_i$ and $b_i$ are independent Gaussian random variables
with mean zero and
$\langle a_i^2 \rangle = \langle b_i^2 \rangle = \frac{2\pi}{T} \times
F(k_i)$. In
the limit $N,T\to\infty$ such that $N/T\to\infty$ 
the correlation function is $g(x)=\int_0^\infty F(k) \cos (kx) dx$.
$F(k)$ is called a spectral density. When $F(k)$ is a wide or narrow
distribution, the process is, respectively, said to be wide- or narrow-band.
The 3D analog is
\begin{equation} 
y (\mbox{\boldmath$r$})= \sum_{l=-N}^{N} \sum_{m=-N}^{N} \sum_{n=-N}^{N}
c_{lmn} e^{i\mbox{\boldmath$k$}_{lmn}.\mbox{\boldmath$r$}}
\label{y3D}
\end{equation}
where
$\mbox{\boldmath$k$}_{lmn}=\frac{2\pi}{T}
(l\mbox{\boldmath$i$}+m\mbox{\boldmath$j$}+n\mbox{\boldmath$k$})$
and $c_{l,m,n}$=$a_{lmn}+ib_{lmn}$.
For $y$ real and $\langle y\rangle=0$ we take
$c_{l,m,n}$=$\bar c_{-l,-m,-n}$ and $c_{0,0,0}$=$0$.
As above, $a_{lmn}$ and $b_{lmn}$ have mean zero and
$\langle a_{lmn}^2
\rangle$=$\langle b_{lmn}^2 \rangle$=$
\frac12 \left({\frac{2\pi}{T}}\right)^3\times\rho(k_{lmn})$. In this case
$g(r)={4\pi}{r^{-1}}\int_0^{\infty}4\pi k \rho (k) \sin kr\; dk$.
A two-dimensional random field is shown in Fig.~\ref{cdf}(a).

A two-phase level-cut GRF model is specified by the  microstructure indicator
function $I({\bf r})=H(\beta-y({\bf r}))$, where $H$ is the Heaviside step
function.
There are two very useful properties of the Gaussian model. First
the random field is ergodic (ensemble averages equal spatial averages),
and second the variables $y({\bf r}_i)$ $i=1,2,3\dots$ and
their spatial derivatives ($\nabla y({\bf r}_i)$,  {\em etc.})
are correlated Gaussian random variables with known
joint probability distribution.
This allows many useful statistical properties of the
thresholded model (such as $S_{11}$) to be calculated.

In the previous section we derived the properties of
the independent interval process in terms
of $p_1$ and $p_2$. However, it is clear that if any two
of the microstructure functions
($p_1$, $p_2$, $S_{11}$, $S_{c1}$, $S_{cc}$) are known
the remainder can be determined by means of
Eqs.~(\ref{s11hat}),~(\ref{sc1hat}) and~(\ref{scchat}).
Since $S_{11}$, $S_{c1}$ and $S_{cc}$ can be calculated for
the level-cut GRF model, this allows an approximation
to be derived for the chord functions. 
The accuracy of the results will then depend on the validity
of the hypothesis that successive chords of the model
are uncorrelated (or nearly so).
For simplicity we consider the case where
$S_{11}$ and $S_{c1}$ are known. Simultaneous solutions
of Eqs.~(\ref{s11hat}) and~(\ref{sc1hat}) then give
\begin{eqnarray}
\hat p_1&=&\frac
{n_c-s(\hat S_{c1}  - s \hat S_{11}+\phi_1 )}
{n_c-s(\hat S_{c1}  + s \hat S_{11}-\phi_1 )} \label{r12eqn1} \\
\hat p_2&=&\frac
{\hat S_{c1}  + s \hat S_{11}-\phi_1 }
{\hat S_{c1}  - s \hat S_{11}+\phi_1 }
\label{r12eqn2}
\end{eqnarray}
This result can be considered a generalization of a recent
approximation developed independently by the authors
of Refs.~\cite{DerridaPers} and~\cite{MajumdarPers}.
In a study of the zero-threshold case
[where $p(z)$=$p_1(z)$=$p_2(z)$] they found
\begin{equation}
\hat p =\frac{n_c  + s ( 2 \hat S_{11}-1 ) }
{n_c  - s ( 2 \hat S_{11}-1 ) }
\label{derrida}
\end{equation}
where $S_{11}(x)=\frac14+\frac 1{2\pi}\arcsin [g(x)]$.
This result is obtained by substituting
$\phi_1=\phi_2=\frac12$ and $S_{c1}=n_c/2$
(which is true for any symmetric medium)
into Eqs.~(\ref{r12eqn1}) and~(\ref{r12eqn2}).
The result provided an excellent approximation
for the case $g(x)=[1/\cosh(x/2)]^{d/2}$ ($d=1,2,3$).

Clearly other approximations can be obtained using the independent
interval process, and several have been previously given. For
example Eq.~(\ref{mcfadden}) was obtained by McFadden~\cite{McFadden56}
and Rice~\cite{Rice58}, and approximate forms of
Eq.~(\ref{MK}) have been used to obtain the
distribution of failure times~\cite{Nielsen90}.  If the chords are
uncorrelated all the methods will give identical
results. An advantage of approximation~(\ref{derrida}) 
and its generalization [Eqs.~(\ref{r12eqn1}) and~(\ref{r12eqn2})] is that
$S_{11}$ and $S_{c1}$ are relatively simple to evaluate;
the functions $R_{uu}$, $R_{dd}$ and
$R_{ud}$ appearing in the integral equations Eq.~(\ref{mcfadden})
and~(\ref{MK}) are quite complex for a non-zero threshold~\cite{Rice58}.

To evaluate the approximations for an arbitrary
threshold we have the following results,
\begin{eqnarray}
\label{pnleqn}
\phi_1 &=& \frac12+\frac12{\rm erf}\frac{\beta}{\sqrt{2}};\;\;\; 
n_c=\frac{\gamma}{\pi}e^{-\frac12 \beta^2} \\
S_{11} &=& \phi_1^2+\frac{1}{2\pi}\int_0^{g(x)} \label{s11grf}
 \frac{dt}{\sqrt{1-t^2}} \exp\left({-\frac{\beta^2}{1+t}}\right)    \\
S_{c1} &=& \frac{n_c}2+ \frac{n_c}2 \nonumber
{\rm erf}\left[\frac{\gamma \beta (1-g) }{\sqrt{2|G|}} \right] \\
&&-\frac{g'\exp\left( -\frac{\beta^2}{1+g}\right)}{2\pi \sqrt{1-g^2}}
{\rm erf} \left[ \frac{\beta}{\sqrt{2|G|}} g' \sqrt{\frac{1-g}{1+g}} \right]
\label{Sc1eqn}
\end{eqnarray}    
where $\gamma=\sqrt{-g''(0)}$
and $|G|=\gamma^2(1-g^2(x))-[g'(x)]^2$. 
The results for $n_c$~\cite{RicePR} and $S_{11}$~\cite{Berk91} are well known,
and the final expression for $S_{c1}$ can be evaluated using
the method of Rice as follows.
For the level-cut Gaussian random field, we have
\begin{equation}
S_{c1}=\langle \delta(\beta-y(x_1)) |y'(x_1)| H(\beta-y(x_2)) \rangle
\end{equation}
The variables ${\bf w}=[w_1,w_2,w_3]=[y(x_1),y(x_2),y'(x_1)]$
have Gaussian distributions with cross
correlation matrix
\begin{equation}
g_{ij}=\langle w_iw_j\rangle\;\; \Rightarrow \;\; G= 
\left[
\begin{array}{ccc} 1 & g(x) & 0 \\ g(x) & 1 & -g'(x) \\
0 & -g'(x) & -g''(0)
\end{array} \right]
\end{equation} for $x=|{x}_2-{x}_1|$ and
${x}_2>{x}_1$. If $x_2<x_1$, $\langle y(x_1)y'(x_2) \rangle=g'(|x|)$
but this not effect the final result.
To find $S_{c1}$ we must therefore evaluate
\begin{equation}
\int\!\!\int\!\!\int\!\! { d {\bf w}}
\delta(\beta-w_1) H(\beta-w_2) |w_3|
\frac{e^{-\frac12 {\bf w}^T G^{-1} {\bf w}} }{(2\pi)^\frac32|G|^\frac12}.
\end{equation}
The integrals extend over all space and the final factor
is just the joint probability density function of $w_i$.
The result is given in Eq.~(\ref{Sc1eqn}).

To obtain $p_i$ it is necessary to invert
Eqs.~(\ref{r12eqn1}) and~(\ref{r12eqn2}).
This can be done using a short and efficient algorithm~\cite{AbateILT}.
As previously noted, $\hat p_i$ needs
to be known to around nine significant figures
to achieve four significant figure accuracy in the result~\cite{QuinTorq96a}.
To minimize cancellation errors in the numerators
and save one integration, we rewrite Eqs.~(\ref{r12eqn1}) and~(\ref{r12eqn2}) as
\begin{eqnarray}
\label{numeric1}
\hat p_1&=&\frac
{\widehat{S_{11}''}-s \widehat{S_{c1}^T} }
{n_c- \widehat{S_{11}''} - s\widehat{S_{c1}^T}  } \\
\hat p_2&=&\frac
{\widehat{S_{11}''}+s \widehat{S_{c1}^T} }
{n_c- \widehat{S_{11}''} + s \widehat{S_{c1}^T}  }
\label{numeric2}
\end{eqnarray}
where $S_{c1}^T(x)=S_{c1}(x)-\frac12 n_c$.
The Laplace transforms $\widehat{S_{11}''}(s)$ and
$\widehat{S_{c1}^T}(s)$ on the right-hand side of
Eqs.~(\ref{numeric1}) and~(\ref{numeric2}) can
be evaluated using numerical quadrature.

To check the validity of the independent interval approximation,
we measure the chord-distribution directly from realizations
of the thresholded model. This is simpler (and minimizes finite-size
effects) in one dimension. A 1D random process $y_1(x)$
can be obtained from a 3D GRF $y_3({\bf r})$ by taking
$y_1(x)$=$y_3({\bf r}_0+\hat {\bf n} x)$ where ${\bf r}_0$ is an arbitrary  
origin and $\hat {\bf n}$ is a unit-vector with arbitrary orientation.
Now $y_1(x)$ can be generated independently of $y_3({\bf r})$ by using
the 1D definition for $y(x)$ given in Eq.~(\ref{y1D}).
To ensure that $y(x)$ and $y_1(x)$ are statistically identical
they must share $g(x)$. This is true if
$F(k)= 4\pi \int_k^\infty s \rho(s) ds$, where $F$ and $\rho$ are,
respectively, the spectral densities of the 1D and 3D random fields.
This shows that $F(k)$ must be a non-decreasing
function for 1D random processes obtained from 3D random
fields.

\begin{figure}
{\samepage\columnwidth20.5pc
\centering \epsfxsize=8.0cm \epsfbox{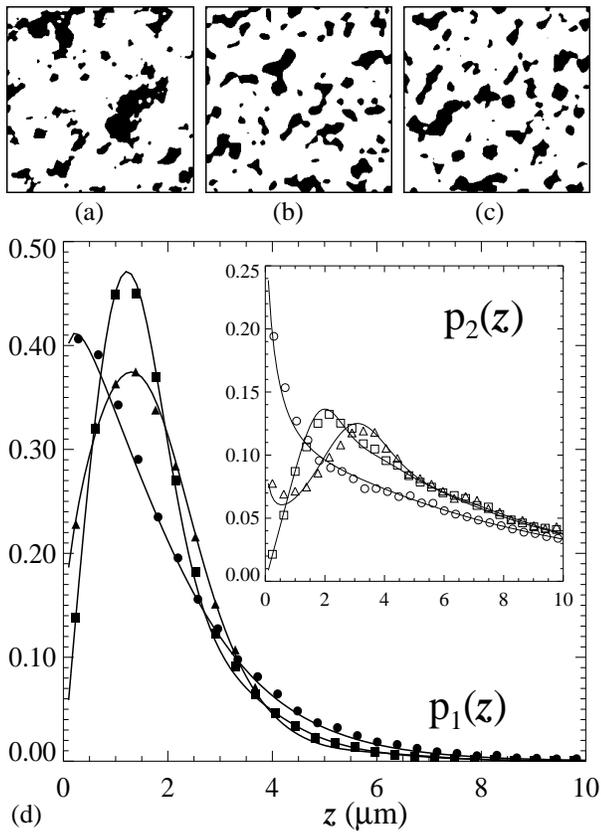} 
\caption{Chord functions of single-cut Gaussian random fields
at volume fraction $\phi_1=0.2$. The symbols are directly measured
from simulations, and the lines correspond to the independent
interval approximation. The models [see Eqs.~(\ref{ga1})--(\ref{gc2})] 
are shown in the top row with side-length 40$\mu$m: 
(a) $g_a(x)$ -- $\circ$; (b) $g_b(x)$ -- $\Box$; (c) $g_c(x)$ -- $\triangle$.
The chord functions correspond to the first-passage time problem
considered by Rice.
\label{2DN0}}
}
\end{figure}

In the modeling of random media, the following Fourier transform pairs
[$g(x)$ and $\rho(k)=-F'(k)/(4\pi k)$] have proved useful
\begin{eqnarray}
\label{ga1}
g_a&=&e^{-x/\xi}(1+\frac x\xi) \\
F_a&=&\frac{4 \xi}{\pi ( 1 + \xi^2 k^2 ) ^2} \label{ga2}\\
\label{gb1}
g_b&=&e^{-x^2/l_0^2} \\
F_b&=&\frac{l_0}{\sqrt \pi} e^{-\frac14 k^2 l_0^2} \label{gb2}
\\
\label{gc1}
g_c&=&e^{-x/\xi}(1+x/\xi) \frac{\sin{2\pi x/d}}{2\pi x/d} \\
F_c&=& \nonumber
\frac{d}{2\pi^2} \left( \tan^{-1} c_- + \tan^{-1} c_+  +
\frac{c_+}{1+c_+^2} \right. \\ && \left. +\frac{c_-}{1+c_-^2} \right)
;\;\;\;\; c_{\pm}=z(\frac{2\pi}{d}\pm k)
\label{gc2}
\end{eqnarray}
For a finite number of crossings per unit length (or specific surface in
three dimensions) it is necessary that $g'(0)=0$~\cite{RicePR}.
For simplicity
we restrict attention to the following parameters which give
$\gamma=\sqrt{-g''(0)}=1\mu$m$^{-1}$:
(a) $\xi=1\mu$m; (b) $l_0=\sqrt{2}\mu$m; and (c)
$\xi= \sqrt{2}\mu$m,  $d=4\pi/\sqrt{6}\mu$m.
A cross-section of the two-phase medium generated in each of the
three cases is shown in Fig.~\ref{2DN0}(a)-(c).
We have checked the approximation in the
volume fraction range $\phi_1\in[0.1,0.9]$; results for $\phi_1$=0.2 are
shown in Fig.~\ref{2DN0} (and are typical of those at other
volume fractions).
The independent interval approximation is seen to provide
remarkably accurate estimates of the measured chord-distributions.

\begin{figure}[bt!]
{\samepage\columnwidth20.5pc
\centering \epsfxsize=8.0cm \epsfbox{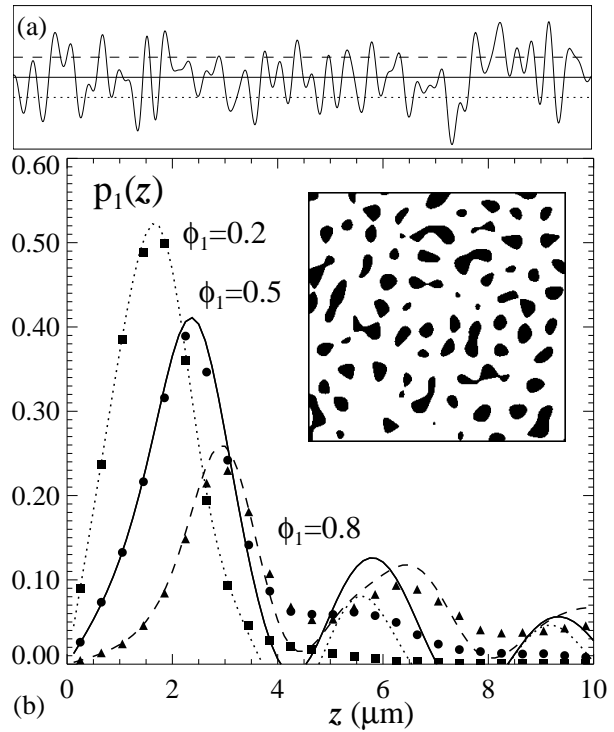}
\caption{The phase 1 chord function of a level-cut GRF with
$g(x)=\sin (kx)/(kx)$ ($k=\sqrt{3}\mu$m) at three different
volume fractions. The inset shows the microstructure at $\phi_1=0.2$
(side-length 40$\mu$m). A 1D transect
(length 150$\mu$m) of the random field is shown in (a). The horizontal
lines corresponds to the threshold at each volume fraction.
The independent interval assumption breaks down because
the process is nearly periodic in some regions.
\label{modSTD}}
}
\end{figure}

\begin{figure}
{\samepage\columnwidth20.5pc
\centering \epsfxsize=8.0cm \epsfbox{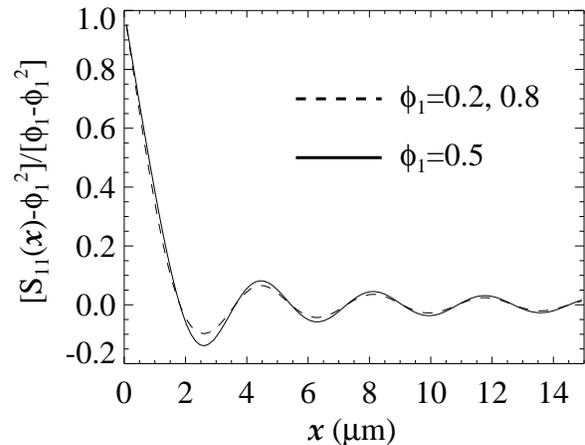}
\caption{The normalized two-point correlation function [Eq.~(\ref{s11grf})]
of the random model shown in Fig.~\protect\ref{modSTD}. 
The strong oscillations in $S_{11}(x)$ correspond to periodic
correlations in the microstructure.
\label{tp2_pall}}
}
\end{figure}

The largest deviations between simulation and the approximation are
seen for the
oscillatory correlation function $g_c(x)$.
We can investigate this ``narrow-band" limit by taking
$\xi\to\infty$ in $g_c(x)$ which gives,
\begin{eqnarray}
g_d(x)&=&\frac{\sin{2\pi x/d}}{2\pi x/d} \\
F_d&=&\frac d{2\pi}H \left( \frac{2\pi}d -k \right).
\end{eqnarray}
The results for $p_1$ at volume fractions $\phi_1$=$0.2,0.5$ and
$0.8$ are shown in Fig.~\ref{modSTD}, and $S_{11}(x)$ is shown
in Fig.~\ref{tp2_pall}.

\noindent
At $\phi_1$=$0.5$ the
approximation is equivalent to that of Refs.~\cite{DerridaPers,MajumdarPers}.
The approximation breaks down after one wave-length $d$ and actually falls
below zero (which is not inconsistent with the derivation).
This is because the process [see Fig.~\ref{modSTD}(a)] has approximately
periodic regions, extending over several wavelengths, which implies some
level of correlation between adjacent chords. 
For example, at $\phi_1$=$0.5$, a chord
of length $\approx\frac12 d$ is more likely
to be followed by another of approximately the
same length than if it were randomly chosen according to the probability
distribution $p_2(z)$. This contradicts the assumptions of the
independent interval approximation. The oscillations in the auto-correlation
function (Fig.~\ref{tp2_pall}) clearly reflect appreciable order in the
system.

Note that $g_d(x)$ represents the worst-case (or most narrow-band) process
corresponding to a 3D field since
$\rho_d(k) \propto \delta ( 2\pi/d -k)$ (ie. an infinitely narrow band
pass filter). However, the related 1D process
corresponds to a low pass filter (ie. it is not strongly narrow-band).


\noindent
This shows that approximations valid for medium and wide-band
Gaussian random processes are sufficient to reproduce the
chord functions of 3D models based on random fields.

The failure of the independent interval approximation in this
narrow-band limit is not critical for two reasons.
First, the model $g_c(x)$ (with $\xi>0$, for which the approximation
is reasonable) has been found more relevant to physical materials than
model $g_d(x)$ (e.g., see Sec.~\ref{applics}).
Second, even in the worst case, the approximation remains
useful out to one wave-length. This may prove adequate for material
characterization.

\section{Extension to more complex models}

\begin{figure}
{\samepage\columnwidth20.5pc
\centering \epsfxsize=8.0cm \epsfbox{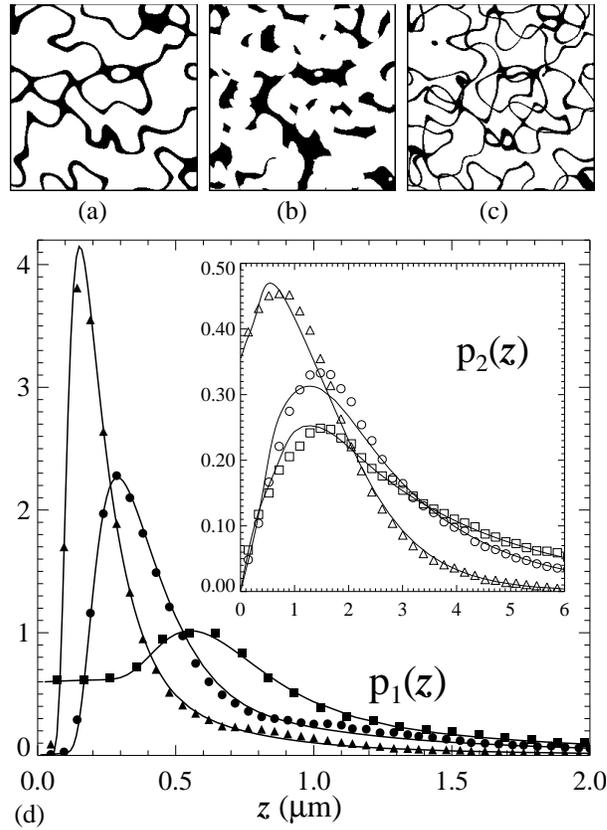} 
\vskip 4mm
\caption{The independent interval approximation compared with
simulations of the chord functions for three distinct models 
based on level-cut Gaussian random fields:
(a) Berk's two-cut model -- $\circ$;
(b) the intersection set of two two-cut models -- $\Box$;
(c) the union set of two two-cut models -- $\triangle$.
The side length of the images is 15$\mu$m.
\label{comp_II}}
}
\end{figure}

So far our results have been concerned with the conventional
first-time distributions associated
with an arbitrary threshold ($\beta$) of a Gaussian random process.
However, the single-cut random field model is not sufficiently
general to model the microstructure of many interesting materials.
To model the bi-continuous structure
of micro-emulsions, Berk~\cite{Berk91} suggested that phase 1 be defined
as the region in space where $\alpha<y({\bf r})<\beta$, this is
the so-called two-cut model [Fig.~\ref{comp_II}(a)]. The two-cut model
has also proven useful in interpreting conductivity
and percolation behavior in polymer blends~\cite{KnackstedtMM1}. Open cell
foams (e.g. aerogels) and the porous network of
sandstones have been
modeled by the intersection sets of two statistically
identical (but independent) two-cut fields [Fig.~\ref{comp_II}(b)],
and closed cell foams may be modeled by the union of
two such structures [Fig.~\ref{comp_II}(c)]~\cite{RobertsRec}. 
Our method can be simply extended to these problems.

For Berk's~\cite{Berk91} two-cut model, we have 
\begin{eqnarray}
\label{Berkphi}
\phi_1 &=& \frac12{\rm erf}\frac{\beta}{\sqrt{2}}-
\frac12{\rm erf}\frac{\alpha}{\sqrt{2}} \\
n_c&=&\frac{\sqrt{\gamma}}{\pi}(e^{-\frac12 \beta^2}+e^{-\frac12 \alpha^2}) \\
\label{Berks11}
S_{11}&=&
\phi_1^2+\frac{1}{2\pi}\int_0^{g(x)}
 \frac{dt}{\sqrt{1-t^2}} \times  \left[
\exp\left({\frac{-\alpha^2}{1+t}}\right) \right.
\label{blank} \nonumber
\\ && \left. 
-2\exp\left({\frac{2\alpha\beta t -\alpha^2
-\beta^2}{2(1-t^2)}}\right)
+\exp\left({\frac{-\beta^2}{1+t}}\right) \right]
\\
\label{Berksc1}
S_{c1} &=& f_{\beta\beta}+f_{\beta\alpha}-f_{\alpha\beta}-
f_{\alpha\alpha}
\end{eqnarray}    
where
\begin{eqnarray}
f_{\alpha\beta} &=& \langle H(\alpha-y(x_1))\delta(\beta-y(x_2)|y'(x_2)| \rangle
\\ \nonumber
&=& \frac{\sqrt{\gamma}e^{-\frac12\beta^2}}{2\pi}
(1+ {\rm erf} \left[\sqrt\frac{\gamma}{2|G|} (\alpha-\beta g)\right] ) \\
&&
-\frac{g'}{2\pi \sqrt{1-g^2}}
\exp\left( -\frac12\frac{\alpha^2-2\alpha\beta g + \beta^2}{1-g^2}
\right) \nonumber \\ &&
\times {\rm erf}
\left[ \frac{\alpha-\beta g}{\sqrt{2|G|}} \frac{g'}{\sqrt{1-g^2}} \right].
\end{eqnarray}

\noindent
Using these results we can directly apply the approximation for the
chord-distributions. We use the field-field function $g_b(x)$
with $l_0=\sqrt{2}\mu$m and consider a ``centered'' two-cut
field ($\alpha=-\beta$) at volume fraction $\phi$=0.2. The results
are shown as circles in Fig.~\ref{comp_II} and show very good agreement with
simulations.

To evaluate the chord functions of the intersection and
union sets we first derive their statistical properties.
Suppose $\Omega({\bf x})$ and $\Psi({\bf x})$ are the indicator functions
of two independent, but statistically identical, models
of random media with properties $\phi_1$, $n_c$, $S_{11}$ and $S_{c1}$.
A new model is obtained by forming the intersection set
of $\Omega$ and $\Psi$ which
has indicator function $I({\bf x})=\Omega({\bf x}) \times \Psi({\bf x})$.
Clearly
$\phi_1^I=\langle I\rangle=\langle\Omega\rangle\langle\Psi\rangle=\phi_1^2$
and $n_c^I=\langle |  I' | \rangle = \langle|\Omega' \Psi +
\Omega \Psi'|\rangle = \langle | \Omega'| \Psi + 
\Omega | \Psi' | \rangle= 2 \langle\Omega\rangle \langle|\Omega'|\rangle=
2\phi_1 n_c$. The relation
$|\Omega' \Psi + \Omega \Psi'|=| \Omega'| \Psi + \Omega | \Psi' |$ is
true everywhere expect where the interface of $\Psi$ and $\Omega$
intersect. The contribution of this error to the
final result is negligible. A similar reasoning can be applied to find
$S_{11}^I$ and $S_{c1}^I$, as well as corresponding results
for an analogously defined union set with indicator function
$ I({\bf x})= \Omega({\bf x})+ \Psi({\bf x})-\Omega({\bf x})
\times \Psi({\bf x})$.
In summary the results needed to apply the approximation for
the chord-distribution function are 
\begin{eqnarray}
\label{SI1a}
\phi_{1}^I&=&(\phi_1)^2 \\  n_c^I&=& 2\phi_1n_c \label{SI1b} \\
\label{SI2a}
S_{11}^I&=&(S_{11})^2 \\
S_{c1}^I&=&2S_{11}S_{c1} \label{SI2b} \\
\label{SU1a}
\phi_{1}^U&=&\phi_1(2-\phi_1) \\ n_c^U&=& 2(1-\phi_1)n_c \label{SU1b} \\
\label{SU2}
S_{11}^U&=&2(\phi_1)^2+2S_{11}(1-2\phi_1)+(S_{11})^2 \\
\label{SU3}
S_{c1}^U&=&2S_{c1}(1-2\phi_1+S_{11})+2n_c(\phi_1-S_{11}).
\end{eqnarray}
Here the un-superscripted microstructure properties correspond
to the primary models $\Omega$ and $\Psi$ and the superscripted
($I$ or $U$) functions are to be used in
Eqs.~(\ref{r12eqn1}) and~(\ref{r12eqn2}).

Although Eqs.~(\ref{SI1a})--(\ref{SU3}) are true for any independent 
random models, we restrict attention to the case where the
primary sets are obtained from Berk's model
[see Eqs.(\ref{Berkphi})--(\ref{Berksc1}) and Fig.~\ref{comp_II}(a)].
As above we consider centered models
($\alpha=-\beta$) at volume fraction $\phi_1$=0.2 obtained from
random fields with correlation function $g_b(x)$ ($l_0=\sqrt{2}\mu$m).
The results of the independent interval approximation
are compared with simulations in Fig.~\ref{comp_II}.
In general the approximation is excellent.  For $r<3\mu$m
significant deviations (up to 10\%) are
seen between the calculated and simulated values of the chord-distribution
of phase 2 ($p_2$). 

\section{Application to porous and composite materials}
\label{applics}

To study the properties of a random medium it is important to have
an accurate model of the microstructure.  If the physical mechanisms
responsible for the evolution of the microstructure are not well known
(or difficult to simulate) an empirically based statistical model
may be useful~\cite{RobertsRec,BourgeoisFilter,YeongRec2,Quiblier84,Adlerbook}.
The level-cut GRF model is well-suited
to this approach because of its generality: the morphology of the
model may be ``tuned" to some degree to match that of the
random medium. The simplest and most common morphological quantities
are the density (or porosity) and the two-point correlation function,
which can both be measured from a cross-sectional image.
It is possible to generate a GRF model
with approximately the same statistical properties by an appropriate 
choice of parameters~\cite{RobertsRec,Quiblier84,Adlerbook}.

\begin{figure}
{\samepage\columnwidth20.5pc
\centering \epsfxsize=8.5cm\epsfbox{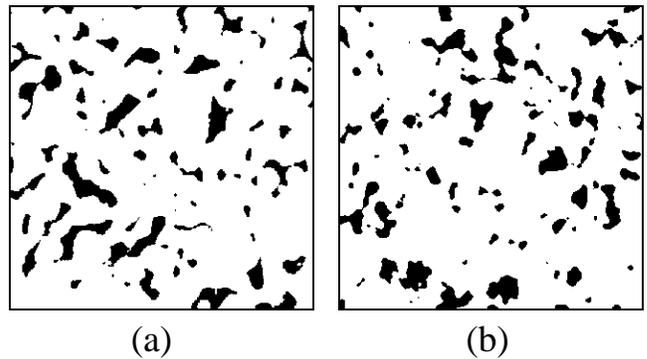}

\vskip 3mm
\caption{A binarized image of a silver-tungsten
composite~\protect\cite{Umekawa65} (a)
compared with a model~\protect\cite{RobertsElasUP} based
on a level-cut Gaussian random field (b). The side-length is
99.4$\mu$m. The parameters and the model are chosen to reproduce the
experimental two-point function and chord-distribution function
(see Fig.~\protect\ref{p2TAg}).
\label{TAg}}
}
\end{figure}

As an example we show a binarized image of a tungsten-silver
composite~\cite{Umekawa65} along side a single-cut GRF model in
Fig.~\ref{TAg}. The parameters of the model were derived in
Ref.~\cite{RobertsElasUP} as follows. The
level-cut parameter is taken as $\beta=-0.84$ so that the
silver volume fraction $\phi_1$=0.2 is exactly that of the
composite [Eq.~(\ref{pnleqn})]. The random-field is
generated using $\rho_c(k)$ [Eq.~(\ref{gc2})]. The length scales of the
random field $\xi=2.15\mu$m and $d=13.0\mu$m are chosen
(by a non-linear least squares

\begin{figure}
{\samepage\columnwidth20.5pc
\centering \epsfxsize=8.5cm \epsfbox{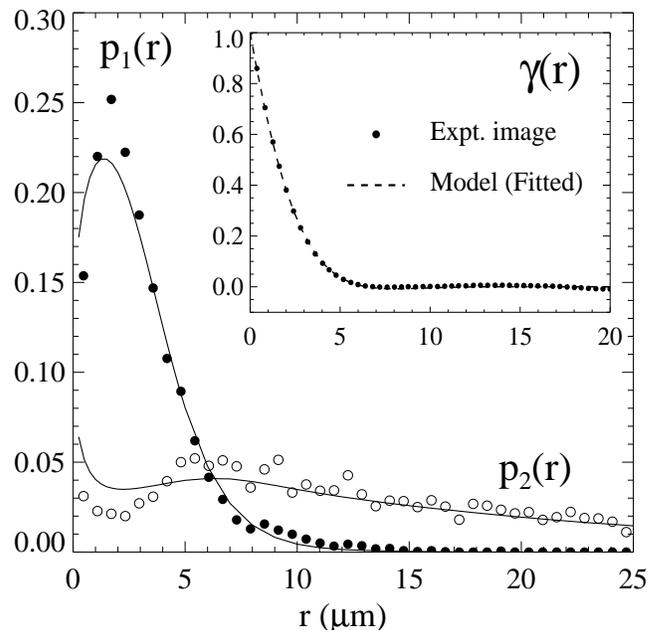}
\caption{The main graph shows a comparison between the chord functions
of a silver-tungsten composite (symbols) and the results of the independent
interval approximation for a level-cut Gaussian random field
model. The composite and model are shown in Fig.~\ref{TAg}. The inset
compares the experimental and model auto-correlation function
$\gamma(x)=(S_{11}(x)-\phi_1^2)/(\phi_1- \phi_1^2)$.
\label{p2TAg}}
}
\end{figure}   

\noindent
method) so that the two point correlation
function of the model matches that of the composite. The theoretical
and experimental values of $S_{11}(r)$ (which are practically
indistinguishable) are shown in the inset of Fig~\ref{p2TAg}.
Since the volume-fraction and two-point function do not uniquely
specify a random
microstructure (ie.\ many different models may reproduce
these morphological quantities~\cite{RobertsRec}), it is necessary
to test the results. The chord functions are ideal in this regard as they
provide a strong signature of microstructure and can be measured
from a cross-sectional image. 
The independent interval approximation and experimental data
are compared in Fig.~\ref{p2TAg}. The reasonable agreement between theory
and experiment indicates that the model is capturing important
features of the tungsten-silver composite.

\begin{figure}
{\samepage\columnwidth20.5pc
\centering \epsfxsize=8.5cm\epsfbox{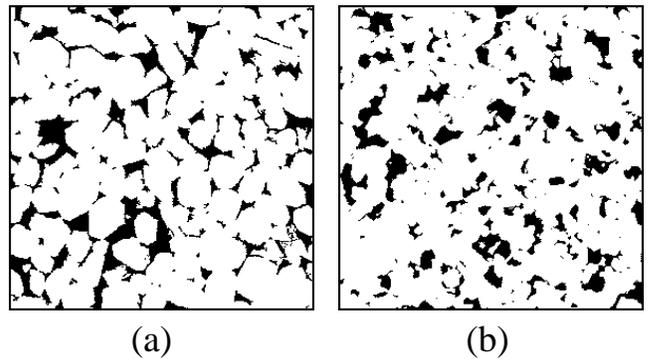}

\vskip 3mm
\caption{A cross-section of Fontainebleau sandstone (a)
compared with a model~\protect\cite{RobertsElasUP} based
on the intersection set of five level-cut Gaussian random fields (b).
The side-length is 2.18mm. The statistical properties of
the sandstone and model are compared in  Fig.~\ref{comp_FB}.
\label{fb1_035}}
}
\end{figure}

\begin{figure}
{\samepage\columnwidth20.5pc
\centering \epsfxsize=8.5cm \epsfbox{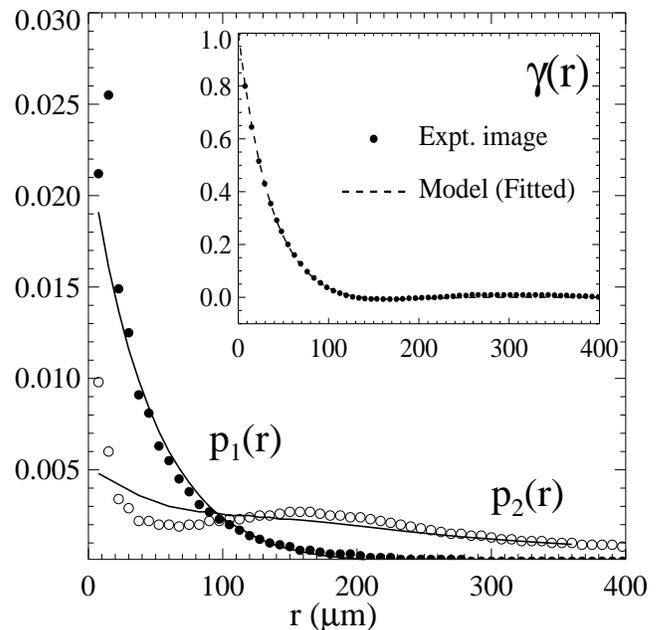}
\caption{The chord functions measured from a 3D
image of Fontainebleau sandstone (symbols) compare well with the results
of the independent interval approximation (main graph) for
a Gaussian random field model.  The inset
shows the the auto-correlation function $\gamma(x)=(S_{11}(x)-\phi_1^2)/(\phi_1-
\phi_1^2)$. Three dimensional realizations of the sandstone and the model
are shown in Figs.~\ref{3Dfb} and~\ref{3DVI5} respectively.
\label{comp_FB}}
}
\end{figure}

\begin{figure}
{\samepage\columnwidth20.5pc
\hskip3.0mm
\centering \epsfxsize=6.5cm \epsfbox{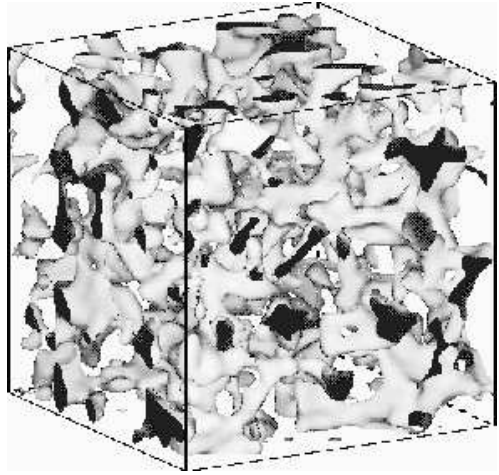}

\vskip 10mm
\caption{3D representation of Fontainebleau sandstone
sample obtained by X-ray tomography~\protect\cite{Schwartz94,Coker96a}.
The pore space is shown as solid to aid visualization. The
side-length of the image is 960$\mu$m. \label{3Dfb}}
}
\end{figure}

\begin{figure}
{\samepage\columnwidth20.5pc
\hskip3.0mm
\centering \epsfxsize=6.5cm \epsfbox{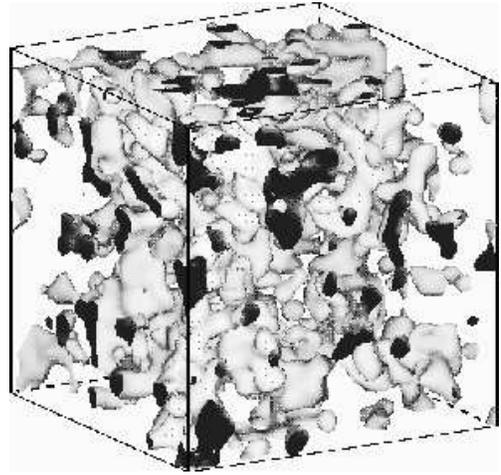}

\vskip 1mm
\caption{A sandstone model based on Gaussian random fields
(cf.\ Fig.~\ref{3Dfb}). The independent interval
approximation for the chord-distribution shows reasonable agreement
with experimental data (Fig.~\ref{comp_FB}). Many of the apparently
isolated regions of pore space (shown as solid) are an artifact of the
plotting procedure.
\label{3DVI5}}
}
\end{figure}

A second example is provided by a digitized image of Fontainebleau
sandstone obtained by
X-ray tomography~\cite{Schwartz94,Coker96a}.
To mimic the granular
character of the sandstone [Fig.~\ref{fb1_035}(a)] we use a model based on
the intersection set of $n$(=5) single-cut Gaussian
random fields. The result is shown in Fig.~\ref{fb1_035}(b).
To match the porosity of the
model with that of the sandstone ($\phi$=0.154), we take  $\beta=0.48$ for
each of the five primary random field models. This corresponds to
$\phi_1=(0.154)^{1/5}$. The experimental two-point function
is reproduced by choosing $\xi=51.9\mu$m and $d=272\mu$m in
model $g_c(r)$ (by a least squares method). The independent interval
approximation for the chord functions is calculated using the relations
$S_{11}^I=(S_{11})^n$ and $S_{c1}^I=n(S_{11})^{n-1} S_{c1}$ [which
are a straightforward extension of Eqs.~(\ref{SI2a}) and~(\ref{SI2b})].
The results are shown in Fig.~\ref{comp_FB}. The model is able to mimic the
the two-point function extremely well, and the chord functions with
good accuracy. This provides evidence that the model
is reasonable. 3D images of the model and sandstone
microstructures are shown in Figs.~\ref{3Dfb} and~\ref{3DVI5}.
The sandstone appears
more well connected than the model; the model showing more isolated
pores. This is actually an artifact of the method used to plot the
pore-solid interface. An algorithm was used to
determine that 98.8\% of the pore-space in the model is
connected to the outer faces which compares well with 99.6\% for
the sandstone.
Therefore the model is also able to capture the interconnections
of the sandstone pores.

\section{Conclusion}
We have derived a semi-analytic approximation
for the chord distribution functions ($p_1$ and $p_2$)
of 3D random media. The approximation is
based on the assumption that successive chord-lengths are uncorrelated.
The result can be applied to models for
which the two-point ($S_{11}$) and 1D ``surface-void''
($S_{c1}$) correlation functions can be evaluated. The calculation
of $S_{11}$ and $S_{c1}$ is generally much
easier than calculation of $p_1$ and $p_2$.
The result is exact for
Boolean models with convex grains since the assumption of
independent intervals is true. We have applied the
approximation to the single level-cut Gaussian random field model
of random materials. In this case the chord functions correspond 
to Rice's first-passage time distribution for random noise.
The approximation is very accurate
for wide-band random fields, but loses accuracy after one ``wavelength''
of the field for extremely narrow band (approximately periodic)
fields.  Note that a narrow-band field corresponds to a low pass
(rather than


\noindent
a narrow pass) filtered process in 1D.

The result also gives accurate results for Berk's two-cut GRF model
and other models based on the intersection and union sets of level-cut
GRF's. This is important for generating 3D models of random
media using empirical information measured from cross-sectional
images; the two-point correlation function does not
necessarily provide sufficient information, and good approximations
for the chord functions are very useful.

\noindent
In this context it is possible to apply the approximation
confidently if the two-point correlation function exhibits
no (or weak) oscillations. 
To demonstrate the application of our results we have compared the
approximation to experimental data obtained from
images of a tungsten-silver composite and a porous sandstone.

In order to derive the chord function approximation we studied the
independent interval process in detail. We have shown that the process
underlies the derivation, and provides useful links between,
important results in many different fields.
The general treatment of the process makes clear
the relation between various approximations for different first-passage times
made in signal theory, the analysis of component failure and persistence
times in coarsening. From the expressions for $\hat p_i$
[Eqs.~(\ref{r12eqn1}) and~(\ref{r12eqn2})] it is simple to obtain the lineal-path functions
as $L_i={\cal L}^{-1}[\phi_i/s-n_c\left( 1+\hat p_i(s)\right) /2s^2]$.
For $i=1$ this is just the ``survival probability'' in the context
of random

\noindent
processes. Similarly the probability density of ``time to failure''
given that $t=0$ falls in a safe region is
$f_1={\cal L}^{-1}[n_c\left(1-\hat p_1(s)\right)/2\phi_1s]$. These expressions 
can be inverted in the same way as $p_i$.
Due to its apparent generality it would be
useful to explore the properties of the process further.
Extensions to include correlation between the chord-lengths and
the development of a 3D analog would be useful
future studies.

\acknowledgements
 
A.P.R.\ gratefully acknowledges the financial support of the
Australian-American Educational Foundation (Fulbright Commission).
S.T. gratefully acknowledges support from the U.S. Department of
Energy, Office of Basic Energy Sciences, under Grant No.\
DE-FG02-92ER14275.


\end{multicols}

\end{document}